 \documentclass[final,5p,times,twocolumn,authoryear]{elsarticle}

\usepackage{amssymb}
\usepackage{fontenc}
\usepackage[utf8]{inputenc}
\usepackage{amsmath}
\usepackage{lipsum}
\usepackage{csquotes}
\usepackage{orcidlink}
\usepackage{ulem}
\usepackage{xcolor}

\journal{Physics Letters B}

\setcitestyle{numbers,square}

\begin{document}

\begin{frontmatter}




\title{Observation of oscillations and near-barrier suppression in the fusion of $^{20}$O + p}

\author[a]{H. Desilets\,\orcidlink{0009-0009-4348-6526}}
\author[a]{Rohit Kumar\,\orcidlink{0000-0002-0450-7218}}
\author[a]{R.~T. deSouza\,\orcidlink{0000-0001-5835-677X}}
\author[a]{S. Hudan\,\orcidlink{0000-0002-9722-2245}}

\author[b]{C. Ciampi\, \orcidlink{0000-0002-8142-0052}}
\author[b]{A. Chbihi\, \orcidlink{0000-0001-5653-4325}}

\author[c]{K.~W. Brown\, \orcidlink{0000-0003-1923-3595}}

\author[d]{Varinderjit Singh\,\orcidlink{0000-0002-9722-2245}}

\author[e]{B. Pinheiro\, \orcidlink{0000-0003-1025-3012}}
\author[e]{J. L. Ferreira\, \orcidlink{0000-0003-2586-8191}}
\author[e]{J. Lubian\, \orcidlink{0000-0001-7735-1835}}

\affiliation[a]{%
Department of Chemistry and Center for Exploration of Energy and Matter, Indiana University
2401 Milo B. Sampson Lane, Bloomington, Indiana 47408, USA}%

\affiliation[b]{GANIL, CEA,DRF-CNRS,IN2P3, 
Blvd. Henri Becquerel, F-14076, Caen, France} %

\affiliation[c]{Facility for Rare Isotope Beams and Department of Chemistry, Michigan State University, East Lansing, MI 48823, USA} %

\affiliation[d]{
Department of Physics, IKG Punjab Technical University, Kapurthala, PB 144603, India} 

\affiliation[e]{%
Instituto de Física, Universidade Federal Fluminense,
Niterói, 24210-340, R.J., Brazil}%

\begin{abstract}
\\
Using an active target detector, the fusion excitation function for $^{20}$O + $^1$H was measured for the first time. Near the barrier, the fusion cross section manifests an oscillatory behavior with broad peaks $\sim$50-100 keV wide. The presence of these peaks likely reflects the low density of low-angular-momentum states in the quasibound regime.  R-matrix coupled channel (CC) calculations that include the first excited 2$^+$ state in $^{20}$O are able to reproduce the observed oscillations. However, one channel CC calculations fail to reproduce the decrease in the sub-barrier cross section experimentally observed. 
\end{abstract}




\end{frontmatter}




\section{Introduction}
\label{introduction}
Nuclear fusion is a fascinating process of both fundamental and practical interest.
While the fusion of the lightest nuclei powers the energy release in stars at their birth, fusion of successively heavier nuclei fuels the later stages of their life cycle \cite{Burbidge57}.
On Earth, synthesis of the heaviest transuranic elements relies on fusion reactions \cite{Oganessian06, Oganessian10}.
Despite significant progress over the last several decades \cite{Back14}, several key questions remain open.
How does the initial structure of the colliding nuclei influence the fusion process? To what extent does the possible formation of transient configurations involving nucleonic clusters \cite{deSouza24} play a role in fusion?
For extremely neutron-rich light nuclei, does polarization of the proton and neutron density distributions \cite{Horowitz08} reflected in low-lying collective modes \cite{Stefanini24} enhance fusion in the sub-barrier regime?

Of particular interest in the fusion of light nuclei is the domain where the level density might be low and not increase monotonically with increasing excitation energy, but exhibits an irregular behavior. A paucity in the density of states has been linked to the suppression of the fusion cross-section at deep sub-barrier energies for $^{12}$C + $^{12}$C \cite{Jiang13, Jiang21, Nippert25}.
An irregular behavior of the level density would mean that the semiclassical description typically used for the fusion of heavy nuclei is no longer valid.

Protons offer a unique probe for exploring fusion in neutron-rich nuclei. 
On general grounds, not only is the Coulomb repulsion between the colliding nuclei minimized as compared to near-barrier fusion of two heavy ions, but the importance of large angular momentum, present in heavy-ion collisions, is also minimized.
Consequently, protons at near-barrier energies provide a sensitive probe of the nuclear potential and how it evolves as a nucleus becomes increasingly neutron-rich. 
Near-barrier proton fusion thus allows exploration of the
limiting case of the fusion process when the number of open channels and many-body states is small. 
To minimize the Coulomb and centrifugal contributions for two heavy ions to a comparable extent, it is necessary to collide the two nuclei at deep sub-barrier energies. Consequently, sensitivity to changes in the nuclear potential for the collision of two heavy ions occurs only at relatively small cross sections. Proton fusion, in contrast, allows one to investigate this behavior for a simpler system at significantly larger cross sections. At near-barrier energies it provides a means to selectively couple to low angular momentum states in the compound nucleus and examine their decay.

\begin{figure}
\begin{center}
\includegraphics[width=0.45\textwidth]{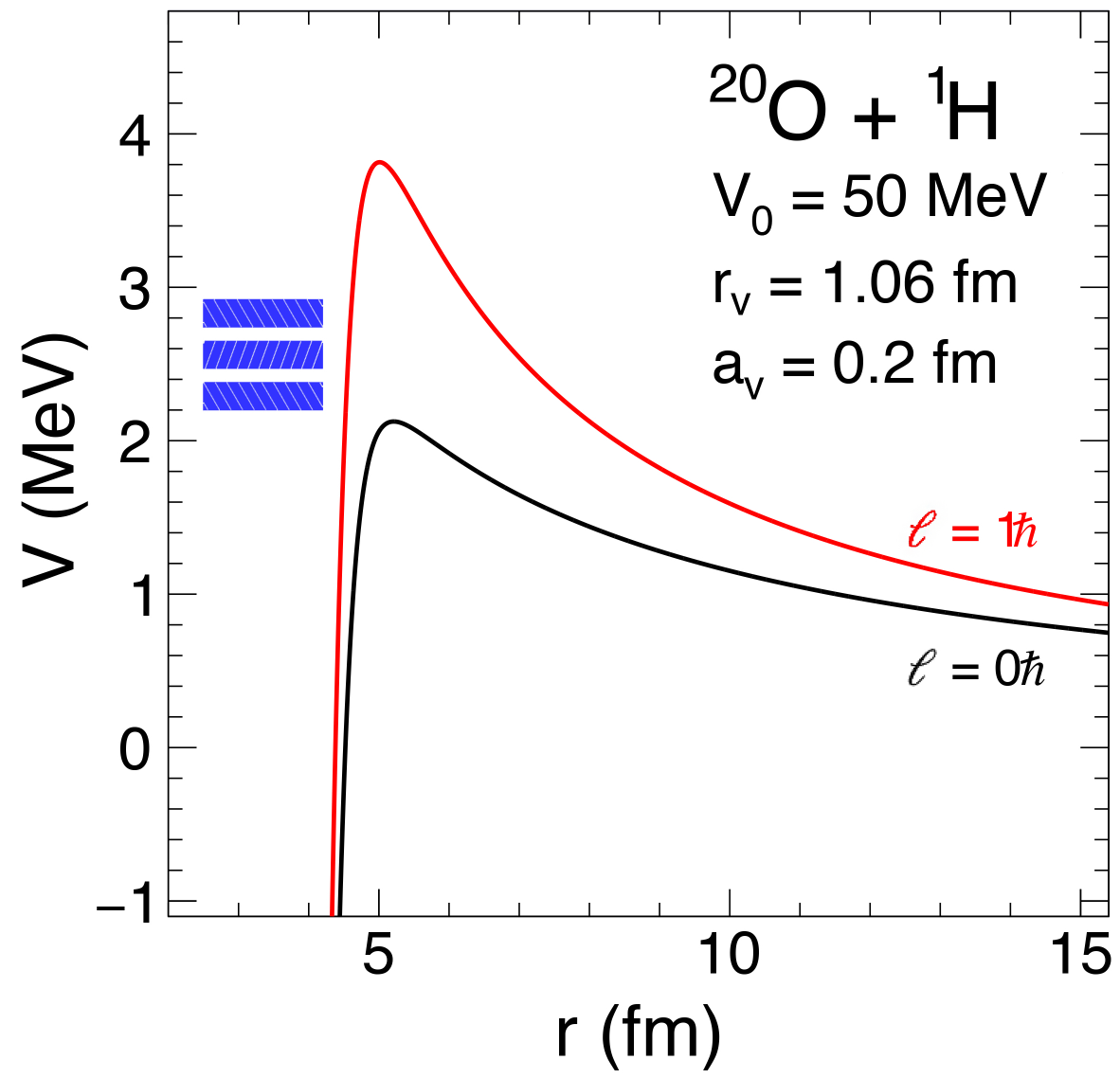}
\caption{Schematic illustration of the potential between a proton and a $^{20}$O nucleus. The short-range attractive part is assumed to have a Woods-Saxon form with the parameters indicated. The cross-hatch regions indicate quasibound states.}
\label{fig:Potential}
\end{center}
\end{figure}

Shown in Fig.~\ref{fig:Potential}, for illustrative purposes, is the one-dimensional potential between an $^{20}$O nucleus and a proton for $\ell$=0$\hbar$ and $\ell$=1$\hbar$. The potential shown is assumed to have a Wood-Saxon form with the indicated values for the parameters. The angular momentum is calculated consistent with a touching spheres configuration. 
Also schematically indicated, by the cross-hatched (blue) regions, are putative short-lived states. These states relative to $\ell$=0$\hbar$ are quasibound states. Observation of such states could indicate not only their existence but also their coupling to the entrance channel, i.e., a large spectroscopic factor. 
In contrast to the potential for heavy ions, the fusion barrier for $\ell$=0$\hbar$ is broad, resulting from the reduced Coulomb interaction between the colliding nuclei and the lack of a centrifugal contribution.
Within this static picture, a proton close to the barrier could sensitively probe: the width of the barrier, the existence of any quasibound states, and the ability of the incoming proton to couple to these states. For reference, the deBroglie wavelength of a 1 MeV proton is $\sim$28 fm, significantly larger than the nuclear diameter of the target $^{20}$O nucleus ($\sim$6 fm). Bombardment with protons at energies just above the barrier essentially provides sensitivity to quasibound states with low angular momentum.
In this work, enabled by the development of radioactive beams and active target detectors \cite{Johnstone21, Carnelli14,Ahn22,Avila17,Asher21b}, we examine the case of fusion of a neutron-rich nucleus, specifically $^{20}$O, with a proton at near-barrier energies. 

\section{Experimental details}
A radioactive beam of $^{20}$O was provided by the SPIRAL1 facility at the GANIL accelerator complex in Caen, France. 
The $^{20}$O (t$_\frac{1}{2}$ = 13.5 s) beam was produced by bombarding a graphite target with a primary beam of $^{22}$Ne at E/A = 80 MeV. The resulting $^{20}$O ions were then accelerated by the CIME cyclotron to an energy of E/A = 2.7 MeV, selected in B$\rho$ by the ALPHA spectrometer, and transported to the experimental setup.
The principal element of the experimental setup was the active-target detector MuSIC@Indiana. Beam impinged on the detector at an intensity up to $\sim$1-2$\times$10$^4$ ions/s. As the MuSIC@Indiana detector has been extensively described in the literature \cite{Johnstone21, Johnstone22, Hudan24, Desilets25, Desilets25a}, only the most salient points are reiterated below.

MuSIC@Indiana is a transverse-field, Frisch-gridded ionization chamber \cite{Johnstone21} in which the anode is segmented into twenty 12.5 mm wide strips oriented transverse to the beam direction. The detector gas in MuSIC@Indiana, CH$_4$ in this experiment, serves as both a target and a detection medium for the analysis. This approach not only provides an energy and angle-integrated measurement of the fusion cross-section at multiple energies simultaneously, but is self-normalized \cite{Carnelli15}.  

In this experiment, MuSIC@Indiana \cite{Desilets25a} was operated at pressure 110 torr $\leq$ P $\leq$ 130 torr. To achieve a 
high resolution scan of the fusion excitation function, the incident beam was also degraded just upstream of the active region \cite{Desilets25}.
The ionization produced as an incident ion traverses the detector is directly measured by the segmented anode.
After being 
processed by high-quality charge-sensitive amplifiers \cite{zepto} and shaping amplifiers the charge deposited on each anode strip
was digitized using peak-sensing ADCs.
All events with incident ions depositing at least 1.0 MeV in the detector were recorded for subsequent analysis. A more detailed description of the design, performance, operation, and calibration of MuSIC@Indiana can be found in \cite{Johnstone21, Johnstone22, Desilets25}.

\section{Experimental Results}

Fusion of $^{20}$O ions with a proton results in the formation of $^{21}$F$^*$ with an excitation energy 12.2 MeV $\leq$ E$^*$ $\leq$ 13.3 MeV due to the Q-value of $\sim$11.1 MeV.
The $^{21}$F$^*$ preferentially decays to $^{20}$F due to the lower neutron binding energy, 8.1 MeV, as compared to that of both a proton or an $\alpha$-particle, 11.1 MeV and 10.3 MeV, respectively.  Thus, by distinguishing F from O ions, measurement of the proton fusion excitation function is accomplished. Calculations with the Hauser-Feshbach statistical decay model code GEMINI++\cite{Charity10} predict that $\geq$99.8$\%$ of the $^{21}$F decays via emission of a single neutron, with the remainder associated with $\alpha$-particle decay. This expectation is borne out by our present measurement in which essentially no $\alpha$-decay is observed. While the channel for a direct $^{20}$O$(p,n)$ reaction is also open (Q$\sim$3 MeV), at the incident energies considered, this cross section is expected to be considerably smaller. 
Calculations in TALYS \cite{Koning23} demonstrate that the $\sigma_{comp}$ is essentially the same as the $\sigma_{rec}$, consistent with the dominance of fusion at the energies measured. 

When an incident ion traverses the detector, the set of measured anode-strip energies, E$_{A}$, is collectively referred to as a \enquote{trace}. The identity of the incident $^{20}$O ion was established by utilizing the
first four anode strips of MuSIC@Indiana for a E$_{A0}$ vs E$_{(A1+A2+A3)}$ measurement as previously reported \cite{Desilets25a}. This selection not only eliminates possible contaminants in the beam but also rejects any reactions originating from the Mylar entrance window of the detector.

To focus on events associated with a reaction in the detector, outgoing $^{20}$O ions were rejected by examining E$_{A18}$ vs E$_{A19}$ and rejecting ions with the beam's characteristic energy deposit. Un-rejected events thus correspond to incident $^{20}$O ions which undergo either reaction or some scattering in the detector. For these remaining events,
proton fusion traces are easily distinguished from traces associated with fusion of $^{20}$O with $^{12}$C due to the much larger specific ionization, thus energy deposit, of the latter case. Proton fusion is also readily distinguished from two-body scattering from either $^{12}$C or protons, which has a characteristic energy deposit in the detector \cite{Johnstone21, Desilets25}. 

\begin{figure}[h]
\begin{center}
\includegraphics[width=0.45\textwidth]{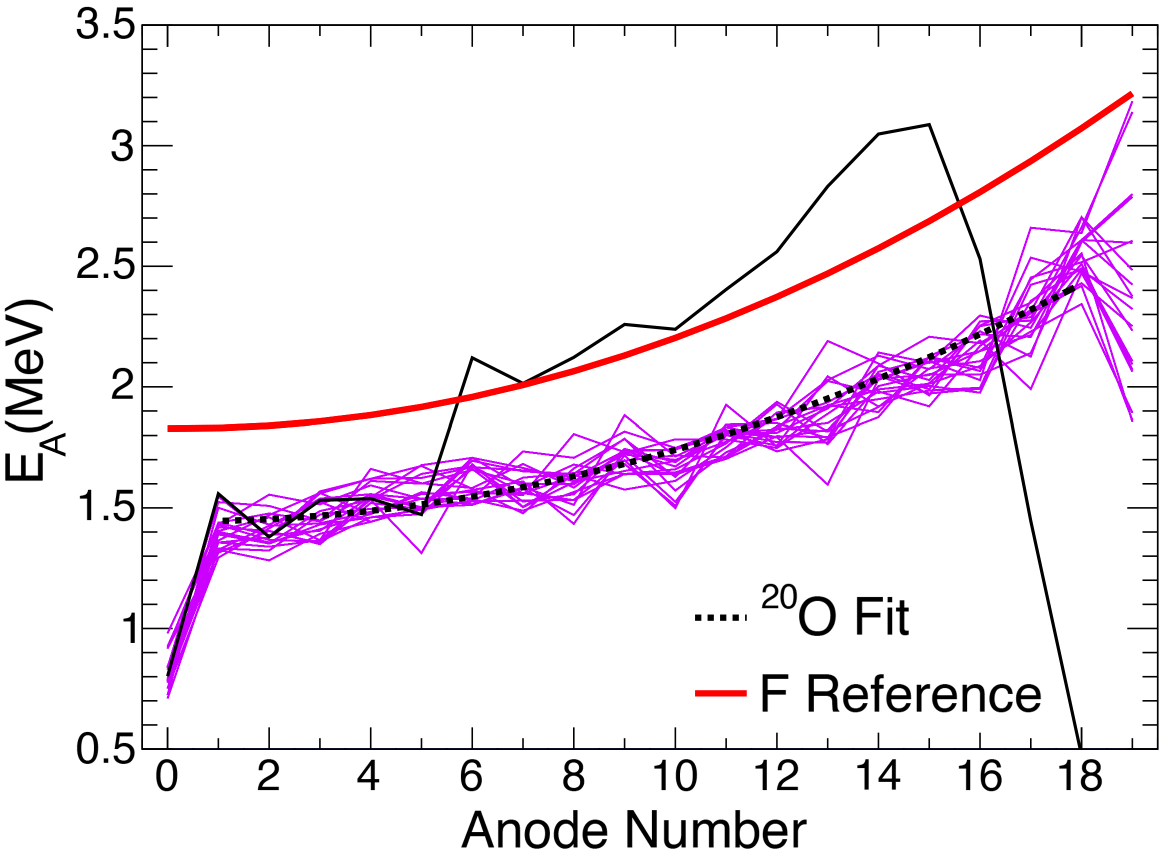}
\caption{Putative proton fusion trace (solid, black) is compared to a few beam traces (solid, magenta). The average of the beam traces used as the reference beam trace is depicted by the dotted (black) line. Based upon a Z$^2$ scaling from the average beam trace, the reference line for F ions is indicated by the solid (red) line.}
\label{fig:Fdeviation}
\end{center}
\end{figure}

Shown in Fig.~\ref{fig:Fdeviation} as solid magenta lines are typical traces for $^{20}$O ions traversing the detector.
One observes that incident $^{20}$O ions produce a characteristic energy deposit on each anode, gradually increasing from $\sim$1.3 MeV upon entering the detector to $\sim$2.4 MeV at the detector exit. A reduced energy deposit is recorded for A0 due to field edge effects. The average energy deposit for all $^{20}$O ions traversing the detector is represented by the dashed black line. This line serves as the reference on which to base the expected energy deposit for a F ion. The F reference line (solid red line) is calculated by scaling the average beam trace by Z$^2$ (i.e., 81/64), consistent with the Bethe-Bloch formula for ions of the same velocity.
In Fig.~\ref{fig:Fdeviation}, one observes a trace (solid black line) that corresponds to only a modest increase in energy deposit relative to the average beam trace (dashed black line) at A6. This modest change is consistent with a small change in atomic number for the ion. This putative F ion corresponds reasonably well to the expected energy deposit for F represented by the solid red line.  For several anodes after its initial increase from beam ionization, the black trace exhibits reasonably good agreement with the F reference.
For this proton fusion event, the $^{20}$F ion reaches a maximum energy deposit in anode 15 before stopping in the detector volume. It can be observed that, as expected, close to the Bragg maximum, the Z$^2$ scaling from the beam reference, without a Bragg peak in the detector, provides a worse description of the fusion event.

\begin{figure}[h]
\begin{center}
\includegraphics[width=0.45\textwidth]{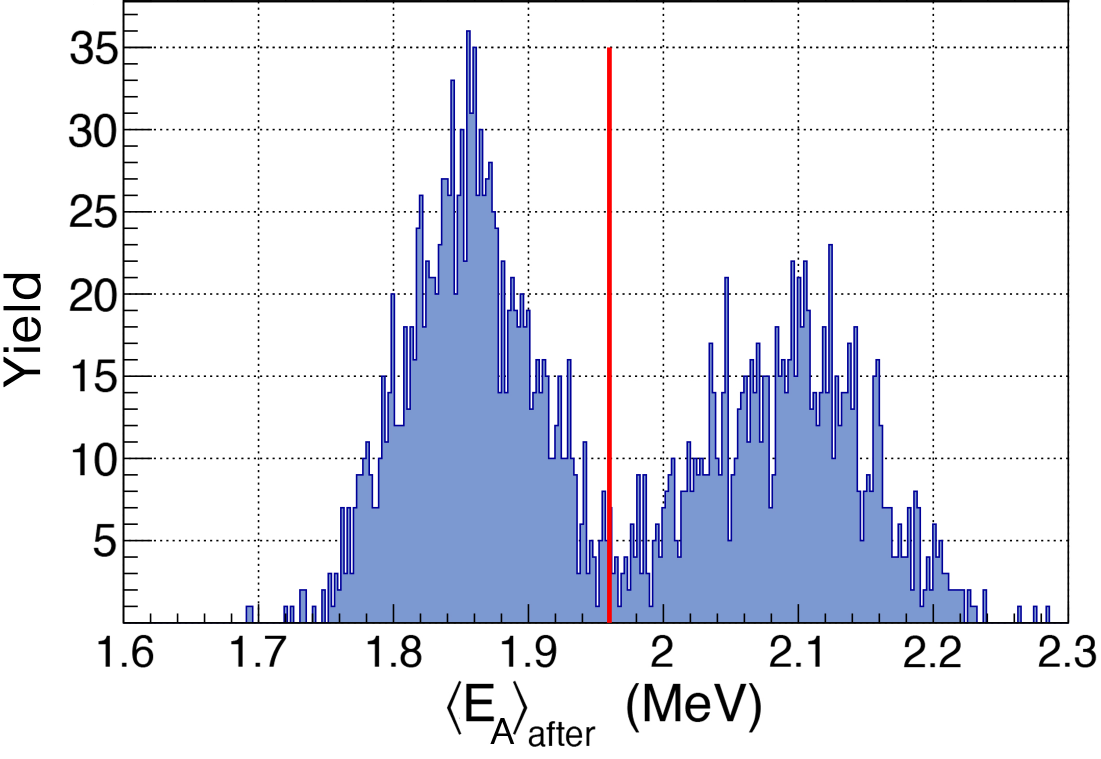}
\caption{For putative proton fusion traces in anode 4, one-dimensional distribution of $\langle$E$\rangle$$_{after}$. Events to the left of the red line are rejected.}
\label{fig:EA}
\end{center}
\end{figure}

The primary means of distinguishing $^{20}$F ions from residual beam is by examining the quantity $\langle$E$\rangle$$_{after}$.
This average energy deposit is constructed from the anode where E$_{A}$ exceeds a threshold energy E$_{th}$ up to the anode where the maximum in E$_{A}$ occurs.
Based upon the energy width of the beam, a threshold energy E$_{th}$ is determined.
E$_{th}$ is taken as 190 keV above the average beam energy for each anode. The distribution of $\langle$E$\rangle$$_{after}$ is presented in Fig.~\ref{fig:EA}.
A bimodal distribution is evident with the higher energy peak in $\langle$E$\rangle$$_{after}$ corresponding to proton fusion and the 
lower energy peak associated with two-body scattering of $^{20}$O ions from a proton.
The solid (red) vertical line indicates the minimum value used to select proton fusion. A similar approach was used by the Argonne group in the analysis of $^{13}$N($\alpha$,p)$^{16}$O \cite{Jayatissa22}.

Determination of the proton fusion cross-section involves counting the number of proton fusion traces relative to the number of incident ions.
Critical in the measurement of the fusion excitation function is the determination of the position at which fusion occurs \cite{Johnstone21, Carnelli14}.
The finite size of each anode strip results in an uncertainty in the energy at which the reaction occurs.  
The fusion cross-section, $\sigma_F$, for each anode segment is related to the number of detected evaporation residues (ERs), $N_{ER}$, by:
$\sigma_F = N_{ER}/(N_{Beam}\times\Delta x)$
where $N_{Beam}$ is the number of incident projectile ions and $\Delta x$ is the thickness of the anode segment. The accuracy of this integrated measurement thus relies simply on the ability to distinguish ionizing events associated with proton fusion from beam or other reactions. Complete detection of the ER within the active volume of MuSIC@Indiana eliminates the need for an efficiency correction. Unlike thin-target measurements, which are sensitive to the measured ER angular distribution,  MuSIC measurements intrinsically provide an angle-integrated measurement of the cross-section. The uncertainty in the measured cross-section is largely determined by the statistical uncertainty associated with $N_{ER}$. In addition, a systematic uncertainty of 5$\%$ associated with the mis-identification of fusion events as non-fusion events for $N_{ER}$ is included.  The uncertainty associated with $N_{Beam}$ is negligible, and the uncertainty associated with $\Delta x$ is defined by the pressure variation in MuSIC@Indiana. All these uncertainties are included in the reported error bars.

\begin{figure}
\begin{center}
\includegraphics[width=0.45\textwidth]{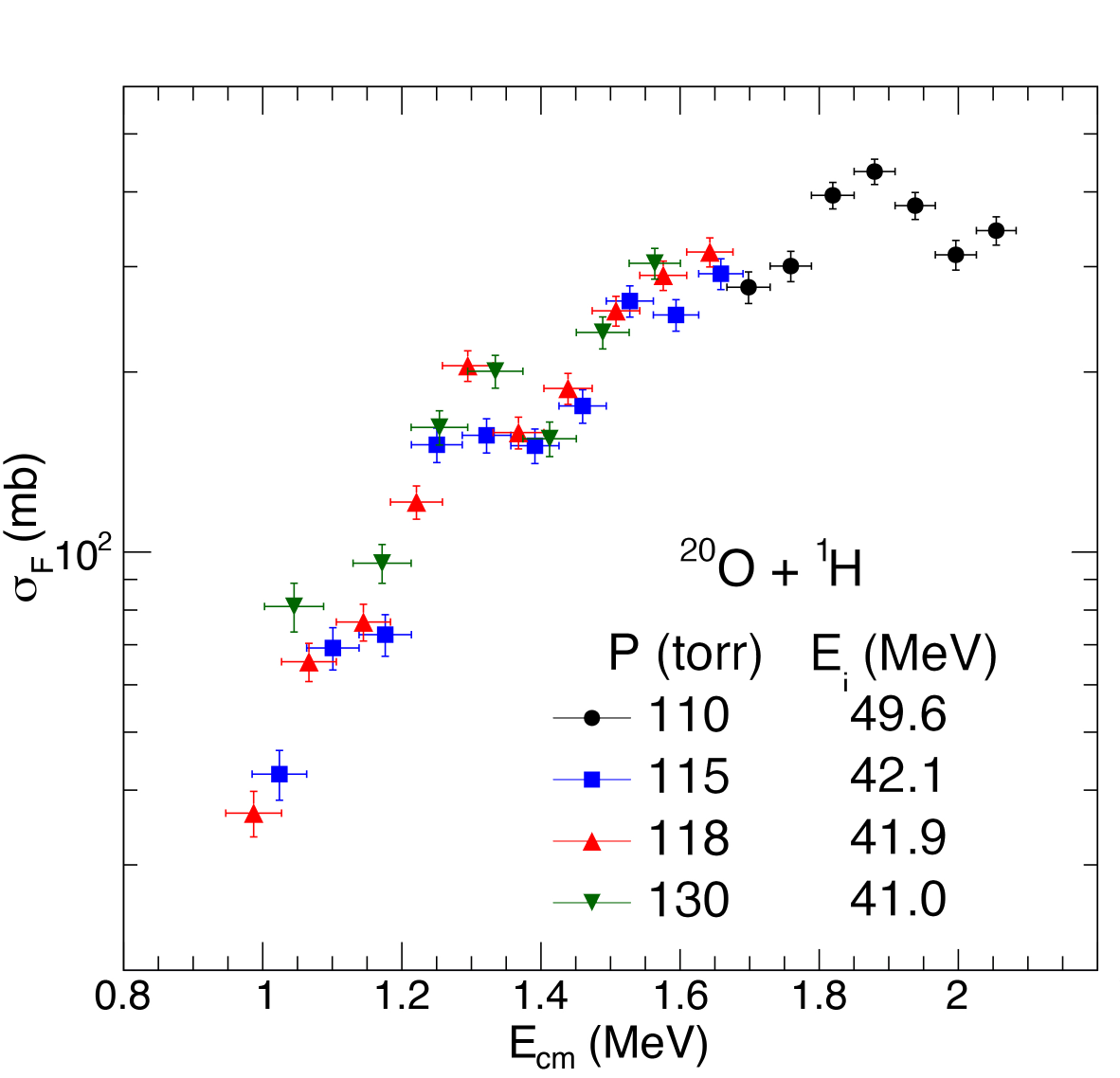}
\caption{Fusion cross section for the reaction $^{20}$O + p. Incident energies at the middle of A0 along with the various pressures used are indicated.}
\label{fig:Xsect}
\end{center}
\end{figure}

Presented in Fig.~\ref{fig:Xsect} is the fusion excitation function for $^{20}$O + p. Indicated by different symbols are the various pressures and incident energies utilized. By varying the pressure and incident energy the fusion excitation is measured over the interval 1.0 $\leq$ E$_{cm}$ $\leq$ 2.05 MeV. Over this interval, one observes a general decrease in the fusion cross-section, $\sigma_F$, indicative of a barrier-governed process.
One also observes that the fusion excitation function is not smooth but exhibits broad peaks superimposed on the overall decrease. These peaks occur at E$_{cm}$ $\approx$ 1.35, 1.55, and 1.85 MeV. If associated with individual quasibound states, the $\sim$50-100 keV width of these peaks indicates a lifetime of $\sim$1-3$\times$10$^{-21}$sec.

These peaks are reminiscent of broad peaks observed for stable nuclei. For an $^{18}$O target, broad peaks were measured in (p,$\gamma$) \cite{Butler59}, (p,n) \cite{Hill56}, and (p,$\alpha$) \cite{Hill56, Clarke59, Carlson61,Lorenz-Wirzba79}. By selecting backward emission, the contribution of direct processes was suppressed, and peaks in the differential cross-section spectrum were described as resonant states in the compound $^{19}$F nucleus.
For the lower energy interval measured in the work of \cite{Lorenz-Wirzba79}, 
the peak centered at E$_p$=846 keV corresponding to an excitation energy E$^*$$\sim$8.8 MeV, had a width
of $\Gamma$=47 keV.
Broad resonances were also observed in proton elastic scattering on $^{18}$O \cite{Yagi62}. The
systematic measurement of proton elastic scattering for the oxygen isotopic chain at near-barrier energies could thus provide a complementary tool to understand the observed oscillatory behavior. Excitation of collective modes in intermediate energy inelastic scattering of protons from $^{18,20,22}$O was attributed to the polarization of the core by the valence neutrons \cite{Chien09}.

In light nuclei, this observation of broad peaks suggests that the standard statistical approach to level density and level properties near the barrier is likely inadequate. In $^{21}$F at an excitation of $\sim$9 MeV, observed individual levels are separated by $\sim$300-600 keV \cite{NuDat3}. At the higher excitation energy of this work, E$^*$$\sim$13 MeV, a separation of $\sim$200 keV is not unreasonable. The observed oscillatory behavior of the fusion cross-section thus suggests that the fusion is occurring in a regime where the level density does not increase monotonically with excitation energy, and the observed peaks correspond to either individual resonances or groups of resonances.

\section{Theoretical Calculations}
An appropriate theoretical tool to describe near-threshold spectroscopy and reactions is the Gamow-shell model in the coupled-channel representation \cite{Wiescher25}. Unfortunately, at the excitation considered, namely E$^*$$\sim$ 11-14 MeV for $^{21}$F, Hamiltonians are at present not sufficiently precise to predict the spin and parity of the resonances above the proton decay threshold.

To provide some insight into the measured fusion excitation coupled channel (CC) calculations were performed. Coupled channel calculations have been successful in describing heavy-ion fusion\cite{Hagino22} with inclusion of the excited states explaining the experimentally measured cross section at sub-barrier energies \cite{Dasso83}. 
These theoretical results for both the one-channel and coupled channel cases were obtained using the FRESCO code \cite{THOMPSON1988167}. The double-folding São Paulo Potential \cite{CPH97} was used as the real part of the optical potential. For the imaginary part, a strong short-range absorptive potential was used. Since the fusion reaction takes place in the inner region of the Coulomb barrier, this potential should be negligible elsewhere. For this reason, we adopted a Woods-Saxon potential with the parameters $\rm{W_{0}=50.0}$~MeV, $\rm{r_{W}=1.06}$~fm, and $\rm{a_{W}=0.2}$~fm \cite{CDH18}.

\begin{figure}
\begin{center}
\includegraphics[width=0.45\textwidth]{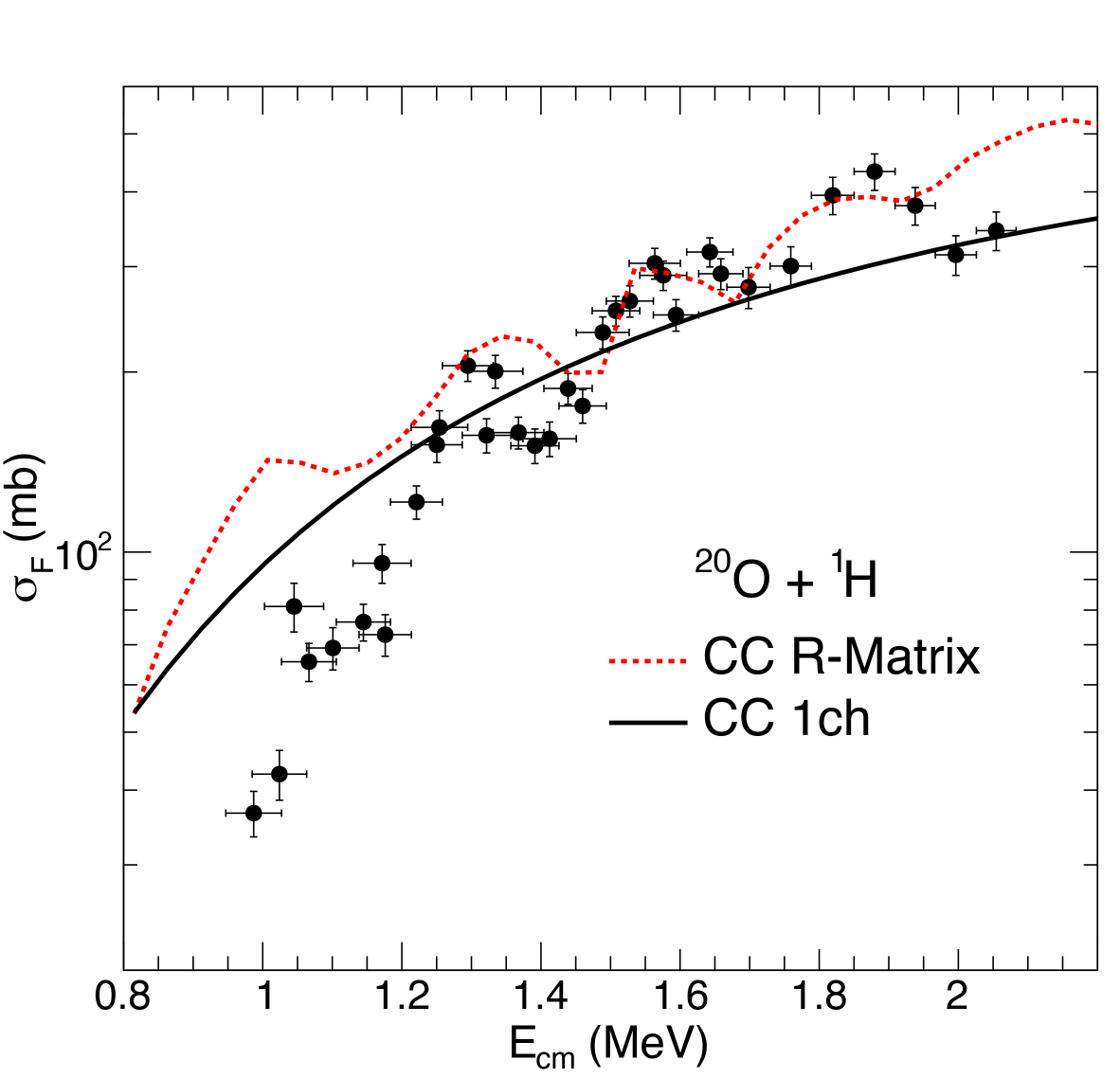}
\caption{Comparison of the experimental fusion cross section with one channel and R-matrix coupled channel calculations.}
\label{fig:XsectwCC}
\end{center}
\end{figure}

The one-channel calculation neglects all channel couplings and solves the Schrödinger equation for the elastic channel. For a spherical potential, this can be written as
\begin{equation}
    \left[\rm T+ V(r)-iW_F(r)\right]\psi(\mathbf{r}) = \rm{E\psi(\mathbf{r})}\text{,}
\end{equation}
where $\rm W_F(r)$ is the short-range potential that takes into account the absorption flux due to the fusion channel. The results can be seen in Fig.\ref{fig:XsectwCC}. It is represented by a black solid curve.
However, it is well-known that the $^{20}$O in its first excited state (2$^{+}$), decays to the ground state by a quadrupole transition, and, therefore, this channel could be important in the reaction dynamics. To account for this excitation, the total wave function of the system was expanded as:
\begin{equation}
    \rm \Psi(\mathbf{r},\xi) = \sum\limits_{\alpha}^{N}\psi_{\alpha}(\mathbf{r})\phi_{\alpha}(\xi)\text{.}
\end{equation}

In the above equation, $\mathrm{N}$ represents the total number of channels that were included in the expansion, labeled $\alpha = 0, 1, 2,\cdots,\rm N$, $\psi_{\alpha}(\mathbf{r})$ is the wave function that describes the relative motion of the system, and $\phi_{\alpha}(\xi)$ is the intrinsic wave function which is an eigenstate of the intrinsic Hamiltonian.
Nonetheless, a channel is open or closed if $\rm E_c$, the threshold energy of a channel c, is smaller or larger than the total energy E of the system. One particularly effective way to solve the CC equations for a closed channel is by using the R-matrix method \cite{Descouvemont_2010}, as closed channels may also appear in the expansion.
When the projectile energy is below the inelastic threshold, those channels are not physically open, but can still contribute virtually. This means that the system can borrow energy virtually, access an inelastic channel, and then return it, modifying the elastic scattering and other reaction channels.

The results of the one-channel and CC R-matrix calculations are presented in Fig.~\ref{fig:XsectwCC} along with the experimental data. The one-channel calculations (solid, black line), i.e., without inclusion of the first excited state (2$^{+}$), do not manifest the oscillatory behavior manifested in the data.   The oscillatory nature of the experimental data is reproduced by the CC R-matrix calculations (dotted, red line).  Remarkably, the centroids of the peaks in the experimental data are also roughly reproduced. This might indicate that the short-lived states are populated by coupling to the 2$^+$ first excited state vibration of the $^{20}$O. Alternatively, it might simply indicate that removal of $\sim$2 MeV (corresponding to the ground-state 2$^+$ energy) from the excited $^{21}$F is necessary to observe the resonant states.

It is also clear that, when comparing the one-channel calculations (solid black line) with the experimental data, the decrease in cross section with decreasing energy is more pronounced in the data than in the calculations. This result can be interpreted as less transmission through the barrier than calculated by the model. Alternatively, it may reflect a low density level compared to statistical expectations, with reduced coupling to the existing states. This observed effect is reminiscent of the suppression observed in deep-sub-barrier fusion with heavier projectiles. For protons, 1 MeV below the Coulomb barrier can be considered deep sub-barrier fusion. Once the proton penetrates the Coulomb barrier, the diabatic approximation is no longer valid; therefore, the interaction potential changes dynamically—a topic that deserves further investigation using time-dependent theories.

\section{Conclusions}
Using an active target approach, fusion of $^{20}$O with a proton was efficiently measured for the first time. Consistent with a barrier-governed process, the fusion cross-section exhibits an overall decrease as the energy decreases. Superimposed on the overall decrease are three broad peaks with a width of $\sim$50-100 keV, indicating that they are short-lived.

One-channel and coupled-channel (R-matrix) calculations were employed to gain a deeper understanding of the measured fusion excitation function. Inclusion of the 2$^+$ first excited $^{20}$O state in the coupled channel calculations reproduced the oscillatory behavior experimentally observed. However, the one-channel calculations overestimated the fusion cross section at the lowest energies measured. 
Proton fusion, with its reduced Coulomb repulsion and minimal centrifugal component as compared to fusion of two heavy ions, provides a particularly effective means to gauge changes in the barrier as one explores even more neutron-rich members of the isotopic chain.

The oscillations observed in the fusion cross section suggest coupling of the low-energy incident proton to specific states in the compound nucleus and their subsequent decay. Unlike other approaches, the present method yields an angle-integrated measurement of these states.
The density of the peaks observed in this work, along with their width, is compatible with the expected (low) resonance density and level width, 1-2 MeV above the proton threshold. The narrow $\ell$-wave distribution, essentially $\ell$=0, thus provides a sensitive probe of low angular momentum quasibound states and their decay. In this domain, the standard ideas typically dominant in heavy-ion reactions are no longer appropriate, and the physics of broad resonances coupled to the continuum is likely more suitable. 
The description of fusion in this quasibound regime and how it evolves into the fusion process of two light heavy-ions such as $^{12}$C+$^{12}$C is an interesting question motivating further work on both experimental and theoretical fronts.

\section{Acknowledgements}
We acknowledge the high-quality beam and experimental support provided by the technical and scientific staff at the Grand Acc\'{e}l\'{e}rateur National d'Ions Lourds (GANIL), in particular D. Allal, B. Jacquot, and D. Gruyer.
We are thankful for the high-quality services of the Mechanical Instrument Services and Electronic Instrument Services facilities at Indiana University.
We thank Marek Ploszajczak, Alex Brown and Kyle Godbey for valuable discussions on the present work.
This work was supported by the U.S. Department of Energy Office of Science under Grant No. 
DE-SC0025230 and Indiana University. This research has received funding from the European Union's HORIZON EUROPE Program under grant agreement no. 101057511. R. deSouza gratefully acknowledges the support of the GANIL Visiting Scientist Program. Kyle Brown acknowledges support from Michigan State University. 
Brazilian authors were supported in part by local funding agencies CNPq, FAPERJ, CAPES, and INCT-FNA (Instituto Nacional de Ci\^{e}ncia e Tecnologia, F\'{i}sica Nuclear e Aplica\c{c}\~{o}es), research Project No. 464898/2014-5.

\textit{Data Availability Statements}. 
The supporting data for this article are from the e831\_{21} experiment and are registered as
https://doi.org/10.26143/GANIL-2023-E831\_21 following the GANIL Data Policy.





\begin{thebibliography}{39}
\expandafter\ifx\csname natexlab\endcsname\relax\def\natexlab#1{#1}\fi
\providecommand{\url}[1]{\texttt{#1}}
\providecommand{\href}[2]{#2}
\providecommand{\path}[1]{#1}
\providecommand{\DOIprefix}{doi:}
\providecommand{\ArXivprefix}{arXiv:}
\providecommand{\URLprefix}{URL: }
\providecommand{\Pubmedprefix}{pmid:}
\providecommand{\doi}[1]{\href{http://dx.doi.org/#1}{\path{#1}}}
\providecommand{\Pubmed}[1]{\href{pmid:#1}{\path{#1}}}
\providecommand{\bibinfo}[2]{#2}
\ifx\xfnm\relax \def\xfnm[#1]{\unskip,\space#1}\fi
\bibitem[{Ahn et~al.(2022)Ahn, Randhawa, Aguilar, Blankstein, Delgado, Dixneuf,
  Henderson, Jackson, Jensen, Jin, Koci, Kolata, Lai, Levano, Li, Mubarak,
  O’Malley, {Ramirez Martin}, Renaud, Serikow, Tollefson, Wilson and
  Yan}]{Ahn22}
\bibinfo{author}{Ahn, T.}, \bibinfo{author}{Randhawa, J.},
  \bibinfo{author}{Aguilar, S.}, \bibinfo{author}{Blankstein, D.},
  \bibinfo{author}{Delgado, L.}, \bibinfo{author}{Dixneuf, N.},
  \bibinfo{author}{Henderson, S.}, \bibinfo{author}{Jackson, W.},
  \bibinfo{author}{Jensen, L.}, \bibinfo{author}{Jin, S.},
  \bibinfo{author}{Koci, J.}, \bibinfo{author}{Kolata, J.},
  \bibinfo{author}{Lai, J.}, \bibinfo{author}{Levano, J.}, \bibinfo{author}{Li,
  X.}, \bibinfo{author}{Mubarak, A.}, \bibinfo{author}{O’Malley, P.},
  \bibinfo{author}{{Ramirez Martin}, S.}, \bibinfo{author}{Renaud, M.},
  \bibinfo{author}{Serikow, M.}, \bibinfo{author}{Tollefson, A.},
  \bibinfo{author}{Wilson, J.}, \bibinfo{author}{Yan, L.},
  \bibinfo{year}{2022}.
\newblock \bibinfo{title}{The notre-dame cube: An active-target time-projection
  chamber for radioactive beam experiments and detector development}.
\newblock \bibinfo{journal}{Nuclear Instruments and Methods in Physics Research
  Section A: Accelerators, Spectrometers, Detectors and Associated Equipment}
  \bibinfo{volume}{1025}, \bibinfo{pages}{166180}.
\newblock \DOIprefix\doi{10.1016/j.nima.2021.166180}.
\bibitem[{Asher et~al.(2021)Asher, Almaraz-Calderon, Baby, Gerken,
  Lopez-Saavedra et~al.}]{Asher21b}
\bibinfo{author}{Asher, B.W.}, \bibinfo{author}{Almaraz-Calderon, S.},
  \bibinfo{author}{Baby, L.T.}, \bibinfo{author}{Gerken, N.},
  \bibinfo{author}{Lopez-Saavedra, E.}, et~al., \bibinfo{year}{2021}.
\newblock \bibinfo{title}{{T}he {E}ncore active target detector: A
  {M}ulti-{S}ampling {I}onization {C}hamber}.
\newblock \bibinfo{journal}{Nucl.Instrum.Meth.A} \bibinfo{volume}{1014},
  \bibinfo{pages}{165724}.
\newblock \DOIprefix\doi{10.1016/j.nima.2021.165724}.
\bibitem[{Avila et~al.(2017)Avila, Rehm, Almaraz-Calderon, Ayangeakaa,
  Dickerson, Hoffman, Jiang, Kay, Lai, Nusair, Pardo, Santiago-Gonzalez, Talwar
  and Ugalde}]{Avila17}
\bibinfo{author}{Avila, M.L.}, \bibinfo{author}{Rehm, K.E.},
  \bibinfo{author}{Almaraz-Calderon, S.}, \bibinfo{author}{Ayangeakaa, A.D.},
  \bibinfo{author}{Dickerson, C.}, \bibinfo{author}{Hoffman, C.R.},
  \bibinfo{author}{Jiang, C.L.}, \bibinfo{author}{Kay, B.P.},
  \bibinfo{author}{Lai, J.}, \bibinfo{author}{Nusair, O.},
  \bibinfo{author}{Pardo, R.C.}, \bibinfo{author}{Santiago-Gonzalez, D.},
  \bibinfo{author}{Talwar, R.}, \bibinfo{author}{Ugalde, C.},
  \bibinfo{year}{2017}.
\newblock \bibinfo{title}{Study of ($\alpha$,p) and ($\alpha$,n) reactions with
  a {M}ulti-{S}ampling {I}onization {C}hamber}.
\newblock \bibinfo{journal}{Nucl. Instr. Meth. A} \bibinfo{volume}{859},
  \bibinfo{pages}{63--68}.
\newblock \DOIprefix\doi{10.1016/j.nima.2017.03.060}.
\bibitem[{Back et~al.(2014)Back, Esbensen, Jiang and Rehm}]{Back14}
\bibinfo{author}{Back, B.B.}, \bibinfo{author}{Esbensen, H.},
  \bibinfo{author}{Jiang, C.L.}, \bibinfo{author}{Rehm, K.E.},
  \bibinfo{year}{2014}.
\newblock \bibinfo{title}{Recent developments in heavy-ion fusion reactions}.
\newblock \bibinfo{journal}{Rev. Mod. Phys.} \bibinfo{volume}{86},
  \bibinfo{pages}{317}.
\newblock \DOIprefix\doi{10.1103/RevModPhys.86.317}.
\bibitem[{Burbidge et~al.(1957)Burbidge, Burbidge, Fowler and
  Hoyle}]{Burbidge57}
\bibinfo{author}{Burbidge, E.M.}, \bibinfo{author}{Burbidge, G.R.},
  \bibinfo{author}{Fowler, W.A.}, \bibinfo{author}{Hoyle, F.},
  \bibinfo{year}{1957}.
\newblock \bibinfo{title}{Synthesis of elements in stars}.
\newblock \bibinfo{journal}{Rev. Mod. Phys.} \bibinfo{volume}{29},
  \bibinfo{pages}{547}.
\newblock \DOIprefix\doi{10.1103/RevModPhys.29.547}.
\bibitem[{Butler and Holmgren(1959)}]{Butler59}
\bibinfo{author}{Butler, J.W.}, \bibinfo{author}{Holmgren, H.D.},
  \bibinfo{year}{1959}.
\newblock \bibinfo{title}{Radiative proton capture by $^{18}$o}.
\newblock \bibinfo{journal}{Phys. Rev.} \bibinfo{volume}{116},
  \bibinfo{pages}{1485}.
\bibitem[{Canto et~al.(2018)Canto, Donangelo, Hussein, Lotti, Lubian and
  Rangel}]{CDH18}
\bibinfo{author}{Canto, L.F.}, \bibinfo{author}{Donangelo, R.},
  \bibinfo{author}{Hussein, M.S.}, \bibinfo{author}{Lotti, P.},
  \bibinfo{author}{Lubian, J.}, \bibinfo{author}{Rangel, J.},
  \bibinfo{year}{2018}.
\newblock \bibinfo{title}{Theoretical considerations about heavy-ion fusion in
  potential scattering}.
\newblock \bibinfo{journal}{Phys. Rev. C} \bibinfo{volume}{98},
  \bibinfo{pages}{044617}.
\bibitem[{Carlson et~al.(1961)Carlson, Kim, Jacobs and Barnard}]{Carlson61}
\bibinfo{author}{Carlson, R.R.}, \bibinfo{author}{Kim, C.C.},
  \bibinfo{author}{Jacobs, J.A.}, \bibinfo{author}{Barnard, A.C.L.},
  \bibinfo{year}{1961}.
\newblock \bibinfo{title}{Elastic scattering and reactions of protons on
  $^{18}$o}.
\newblock \bibinfo{journal}{Phys. Rev.} \bibinfo{volume}{122},
  \bibinfo{pages}{607}.
\bibitem[{Carnelli et~al.(2014)}]{Carnelli14}
\bibinfo{author}{Carnelli, P.F.F.}, et~al., \bibinfo{year}{2014}.
\newblock \bibinfo{title}{{M}easurement of {F}usion {R}eactions of
  {L}ow-{I}ntensity {R}adioactive {C}arbon {B}eams on $^{12}${C} and their
  {I}mplications for the {U}nderstanding of {X}-{R}ay {B}ursts}.
\newblock \bibinfo{journal}{Phys. Rev. Lett.} \bibinfo{volume}{112},
  \bibinfo{pages}{192701}.
\newblock \DOIprefix\doi{10.1103/PhysRevLett.112.192701}.
\bibitem[{Carnelli et~al.(2015)}]{Carnelli15}
\bibinfo{author}{Carnelli, P.F.F.}, et~al., \bibinfo{year}{2015}.
\newblock \bibinfo{title}{{M}ulti-{S}ampling {I}onization {C}hamber ({MUSIC})
  for measurements of fusion reactions with radioactive beams}.
\newblock \bibinfo{journal}{Nucl. Instr. Meth. A} \bibinfo{volume}{799},
  \bibinfo{pages}{197}.
\newblock \DOIprefix\doi{10.1016/j.nima.2015.07.030}.
\bibitem[{Chamon et~al.(1997)Chamon, Pereira, Hussein, C\^andido~Ribeiro and
  Galetti}]{CPH97}
\bibinfo{author}{Chamon, L.C.}, \bibinfo{author}{Pereira, D.},
  \bibinfo{author}{Hussein, M.S.}, \bibinfo{author}{C\^andido~Ribeiro, M.A.},
  \bibinfo{author}{Galetti, D.}, \bibinfo{year}{1997}.
\newblock \bibinfo{title}{Nonlocal description of the nucleus-nucleus
  interaction}.
\newblock \bibinfo{journal}{Phys. Rev. Lett. C} \bibinfo{volume}{79},
  \bibinfo{pages}{5218--5221}.
\newblock \DOIprefix\doi{10.1103/PhysRevLett.79.5218}.
\bibitem[{Charity(2010)}]{Charity10}
\bibinfo{author}{Charity, R.}, \bibinfo{year}{2010}.
\newblock \bibinfo{title}{Systematic description of evaporation spectra for
  light and heavy compound nuclei}.
\newblock \bibinfo{journal}{Phys. Rev. C} \bibinfo{volume}{82},
  \bibinfo{pages}{014610}.
\newblock \DOIprefix\doi{10.1103/PhysRevC.82.014610}.
\bibitem[{Chien and Khoa(2009)}]{Chien09}
\bibinfo{author}{Chien, N.D.}, \bibinfo{author}{Khoa, D.T.},
  \bibinfo{year}{2009}.
\newblock \bibinfo{title}{Neutron transition strengths of 2$_1$$^+$ states in
  the neutron-rich oxygen isotopes determined from inelastic proton
  scattering}.
\newblock \bibinfo{journal}{Phys. Rev. C} \bibinfo{volume}{79},
  \bibinfo{pages}{034314}.
\newblock \DOIprefix\doi{10.1103/PhysRevC.79.034314}.
\bibitem[{Clarke et~al.(1959)Clarke, Almqvist and Paul}]{Clarke59}
\bibinfo{author}{Clarke, R.L.}, \bibinfo{author}{Almqvist, E.},
  \bibinfo{author}{Paul, E.B.}, \bibinfo{year}{1959}.
\newblock \bibinfo{title}{{Properties of levels excited in (p, $\alpha$)
  reactions on $^{18}${O}, $^{31}${P}, $^{35}${C}l, $^{37}${C}l, $^{39}${K} and
  $^{41}${K}}}.
\newblock \bibinfo{journal}{Nucl. Phys.} \bibinfo{volume}{14},
  \bibinfo{pages}{472}.
\bibitem[{Dasso et~al.(1983)Dasso, Landowne and Winther}]{Dasso83}
\bibinfo{author}{Dasso, C.H.}, \bibinfo{author}{Landowne, S.},
  \bibinfo{author}{Winther, A.}, \bibinfo{year}{1983}.
\newblock \bibinfo{title}{Channel-coupling effects in heavy-ion fusion
  reactions}.
\newblock \bibinfo{journal}{Nucl. Phys. A} \bibinfo{volume}{405},
  \bibinfo{pages}{381--396}.
\newblock \DOIprefix\doi{10.1016/0375-9474(83)90578-x}.
\bibitem[{Descouvemont and Baye(2010)}]{Descouvemont_2010}
\bibinfo{author}{Descouvemont, P.}, \bibinfo{author}{Baye, D.},
  \bibinfo{year}{2010}.
\newblock \bibinfo{title}{The {R}-matrix theory}.
\newblock \bibinfo{journal}{Reports on Progress in Physics}
  \bibinfo{volume}{73}, \bibinfo{pages}{036301}.
\newblock \DOIprefix\doi{10.1088/0034-4885/73/3/036301}.
\bibitem[{Desilets et~al.(2025a)Desilets, Kumar and deSouza}]{Desilets25}
\bibinfo{author}{Desilets, H.}, \bibinfo{author}{Kumar, R.},
  \bibinfo{author}{deSouza, R.T.}, \bibinfo{year}{2025}a.
\newblock \bibinfo{title}{Measuring near- and sub-barrier fusion using
  radioactive beams with a {M}u{SIC} detector}.
\newblock \bibinfo{journal}{Nucl. Instr. Meth. in Phys. Res. A}
  \bibinfo{volume}{1075}, \bibinfo{pages}{170440}.
\newblock \DOIprefix\doi{10.1016/j.nima.2025.170440}.
\bibitem[{Desilets et~al.(2025b)Desilets, Kumar, deSouza, Hudan, Ciampi,
  Chbihi, Brown, Godbey, Pinheiro, Cardozo and Lubian}]{Desilets25a}
\bibinfo{author}{Desilets, H.}, \bibinfo{author}{Kumar, R.},
  \bibinfo{author}{deSouza, R.T.}, \bibinfo{author}{Hudan, S.},
  \bibinfo{author}{Ciampi, C.}, \bibinfo{author}{Chbihi, A.},
  \bibinfo{author}{Brown, K.W.}, \bibinfo{author}{Godbey, K.},
  \bibinfo{author}{Pinheiro, B.}, \bibinfo{author}{Cardozo, E.N.},
  \bibinfo{author}{Lubian, J.}, \bibinfo{year}{2025}b.
\newblock \bibinfo{title}{Impact of pairing and neutron-excess on suppression
  of the above-barrier fusion cross-section in $^{19}${O}+$^{12}${C}}.
\newblock \bibinfo{journal}{Phys. Lett. B} \bibinfo{volume}{868},
  \bibinfo{pages}{139643}.
\newblock \DOIprefix\doi{10.1016/j.physletb.2025.139643}.
\bibitem[{deSouza()}]{zepto}
\bibinfo{author}{deSouza, R.T.}, .
\newblock \bibinfo{title}{{Z}epto {S}ystems {I}nc.}
\bibitem[{deSouza et~al.(2024)deSouza, Godbey, Hudan and
  Nazarewicz}]{deSouza24}
\bibinfo{author}{deSouza, R.T.}, \bibinfo{author}{Godbey, K.},
  \bibinfo{author}{Hudan, S.}, \bibinfo{author}{Nazarewicz, W.},
  \bibinfo{year}{2024}.
\newblock \bibinfo{title}{Search for beyond-mean-field signatures in heavy-ion
  fusion reactions}.
\newblock \bibinfo{journal}{Phys. Rev. C} \bibinfo{volume}{109},
  \bibinfo{pages}{L041601}.
\newblock \DOIprefix\doi{10.1103/PhysRevC.109.L041601}.
\bibitem[{H.~Lorenz-Wirzba et~al.(1979)H.~Lorenz-Wirzba, Trautvetter, Wiescher
  and Rolfs}]{Lorenz-Wirzba79}
\bibinfo{author}{H.~Lorenz-Wirzba, P.S.}, \bibinfo{author}{Trautvetter, H.P.},
  \bibinfo{author}{Wiescher, M.}, \bibinfo{author}{Rolfs, C.},
  \bibinfo{year}{1979}.
\newblock \bibinfo{title}{The $^{18}${O}(p, $\alpha$)$^{15}${N} reaction at
  stellar energies}.
\newblock \bibinfo{journal}{Nucl. Phys. A} \bibinfo{volume}{313},
  \bibinfo{pages}{346}.
\bibitem[{Hagino et~al.(2022)Hagino, Ogata and Moro}]{Hagino22}
\bibinfo{author}{Hagino, K.}, \bibinfo{author}{Ogata, K.},
  \bibinfo{author}{Moro, A.M.}, \bibinfo{year}{2022}.
\newblock \bibinfo{title}{Coupled-channels calculations for nuclear reactions:
  From exotic nuclei to superheavy elements}.
\newblock \bibinfo{journal}{Progress in Particle and Nuclear Physics}
  \bibinfo{volume}{125}, \bibinfo{pages}{103951}.
\newblock \DOIprefix\doi{10.1016/j.ppnp.2051}.
\bibitem[{Hill and Blair(1956)}]{Hill56}
\bibinfo{author}{Hill, H.A.}, \bibinfo{author}{Blair, J.M.},
  \bibinfo{year}{1956}.
\newblock \bibinfo{title}{{Y}ields of the $^{18}$o(p, $\alpha$)$^{15}$n and
  $^{18}$o(p, n)$^{18}$f {R}eactions for {P}rotons of 800 kev to 3500 kev}.
\newblock \bibinfo{journal}{Phys. Rev.} \bibinfo{volume}{104},
  \bibinfo{pages}{198}.
\bibitem[{Horowitz et~al.(2008)Horowitz, Dussan and Berry}]{Horowitz08}
\bibinfo{author}{Horowitz, C.J.}, \bibinfo{author}{Dussan, H.},
  \bibinfo{author}{Berry, D.K.}, \bibinfo{year}{2008}.
\newblock \bibinfo{title}{{F}usion of neutron-rich oxygen isotopes in the crust
  of accreting neutron stars}.
\newblock \bibinfo{journal}{Phys. Rev. C} \bibinfo{volume}{77},
  \bibinfo{pages}{045807}.
\newblock \DOIprefix\doi{10.1103/PhysRevC.77.045807}.
\bibitem[{Hudan et~al.(2024)Hudan, Desilets, Kumar, deSouza, Ciampi, Chbihi and
  Brown}]{Hudan24}
\bibinfo{author}{Hudan, S.}, \bibinfo{author}{Desilets, H.},
  \bibinfo{author}{Kumar, R.}, \bibinfo{author}{deSouza, R.T.},
  \bibinfo{author}{Ciampi, C.}, \bibinfo{author}{Chbihi, A.},
  \bibinfo{author}{Brown, K.W.}, \bibinfo{year}{2024}.
\newblock \bibinfo{title}{Influence of additional neutrons on the fusion cross
  section beyond the {N}=8 shell}.
\newblock \bibinfo{journal}{Phys. Rev. C} \bibinfo{volume}{109},
  \bibinfo{pages}{L011601}.
\newblock \DOIprefix\doi{10.1103/PhysRevC.109.L011601}.
\bibitem[{Jayatissa et~al.(2022)}]{Jayatissa22}
\bibinfo{author}{Jayatissa, H.}, et~al., \bibinfo{year}{2022}.
\newblock \bibinfo{title}{First direct measurement of the
  $^{13}${N}($\alpha$,p)$^{16}${O} reaction relevant for core-collapse
  supernovae nucleosynthesis}.
\newblock \bibinfo{journal}{Phys. Rev. C} \bibinfo{volume}{105},
  \bibinfo{pages}{L042802}.
\newblock \DOIprefix\doi{10.1103/PhysRevC.105.L042802}.
\bibitem[{Jiang et~al.(2013)Jiang, Back, Esbensen, Janssens, Rehm and
  Charity}]{Jiang13}
\bibinfo{author}{Jiang, C.L.}, \bibinfo{author}{Back, B.B.},
  \bibinfo{author}{Esbensen, H.}, \bibinfo{author}{Janssens, R.V.F.},
  \bibinfo{author}{Rehm, K.E.}, \bibinfo{author}{Charity, R.J.},
  \bibinfo{year}{2013}.
\newblock \bibinfo{title}{{O}rigin and {C}onsequences of $^{12}${C} +
  $^{12}${C} {F}usion {R}esonances at {D}eep {S}ub-barrier {E}nergies}.
\newblock \bibinfo{journal}{Phys. Rev. Lett.} \bibinfo{volume}{110},
  \bibinfo{pages}{072701}.
\newblock \DOIprefix\doi{10.1103/PhysRevLett.110.072701}.
\bibitem[{Jiang et~al.(2021)Jiang, Back, Rehm, Hagino, Montaganoli and
  Stefanini}]{Jiang21}
\bibinfo{author}{Jiang, C.L.}, \bibinfo{author}{Back, B.B.},
  \bibinfo{author}{Rehm, K.E.}, \bibinfo{author}{Hagino, K.},
  \bibinfo{author}{Montaganoli, G.}, \bibinfo{author}{Stefanini, A.M.},
  \bibinfo{year}{2021}.
\newblock \bibinfo{title}{Heavy-ion fusion reactions at extreme sub-barrier
  energies}.
\newblock \bibinfo{journal}{Eur. Phys. J. A} \bibinfo{volume}{57},
  \bibinfo{pages}{235}.
\newblock \DOIprefix\doi{10.1140/epja/s10050-021-00536-2}.
\bibitem[{Johnstone et~al.(2021)}]{Johnstone21}
\bibinfo{author}{Johnstone, J.E.}, et~al., \bibinfo{year}{2021}.
\newblock \bibinfo{title}{{M}u{SIC}@{I}ndiana: an effective tool for accurate
  measurement of fusion with low-intensity radioactive beams}.
\newblock \bibinfo{journal}{Nucl. Instr. Meth. A} \bibinfo{volume}{1014},
  \bibinfo{pages}{166697}.
\newblock \DOIprefix\doi{10.1016/j.nima.2021.165697}.
\bibitem[{Johnstone et~al.(2022)}]{Johnstone22}
\bibinfo{author}{Johnstone, J.E.}, et~al., \bibinfo{year}{2022}.
\newblock \bibinfo{title}{Improving the characterization of fusion in a
  {M}u{SIC} detector by spatial localization}.
\newblock \bibinfo{journal}{Nucl. Instr. Meth. A} \bibinfo{volume}{1025},
  \bibinfo{pages}{166212}.
\newblock \DOIprefix\doi{10.1016/j.nima.2021.166212}.
\bibitem[{Koning et~al.(2023)Koning, Hilaire and Goriely}]{Koning23}
\bibinfo{author}{Koning, A.J.}, \bibinfo{author}{Hilaire, S.},
  \bibinfo{author}{Goriely, S.}, \bibinfo{year}{2023}.
\newblock \bibinfo{title}{{TALYS}: modeling of nuclear reactions}.
\newblock \bibinfo{journal}{Eur. Phys. J. A} \bibinfo{volume}{59},
  \bibinfo{pages}{131}.
\newblock \DOIprefix\doi{10.1140/epja/s10050-023-01034-3}.
\bibitem[{Nippert et~al.(2025)}]{Nippert25}
\bibinfo{author}{Nippert, J.}, et~al., \bibinfo{year}{2025}.
\newblock \bibinfo{title}{Refining the deep sub-barrier $^{12}${C} + $^{12}${C}
  fusion excitation function with the stella apparatus}.
\newblock \bibinfo{journal}{Phys. Rev. C} \bibinfo{volume}{111},
  \bibinfo{pages}{065804}.
\newblock \DOIprefix\doi{10.1103/PhysRevC.111.065804}.
\bibitem[{NuDat3()}]{NuDat3}
\bibinfo{author}{NuDat3}, .
\newblock \URLprefix \url{http://www.nndc.bnl.gov/nudat3}.
  \bibinfo{note}{{h}ttp://www.nndc.bnl.gov/nudat3}.
\bibitem[{Oganessian et~al.(2006)Oganessian, Utyonkov, Lobanov, Abdullin,
  Polyakov, Sagaidak, Shirokovsky, Tsyganov, Voinov, Gulbekian, Bogomolov,
  Gikal, Mezentsev, Iliev, Subbotin, Sukhov, Subotic, Zagrebaev, Vostokin,
  Itkis, Moody, Patin, Shaughnessy, Stoyer, Stoyer, Wilk, Kenneally, Landrum,
  Wild and Lougheed}]{Oganessian06}
\bibinfo{author}{Oganessian, Y.T.}, \bibinfo{author}{Utyonkov, V.K.},
  \bibinfo{author}{Lobanov, Y.V.}, \bibinfo{author}{Abdullin, F.S.},
  \bibinfo{author}{Polyakov, A.N.}, \bibinfo{author}{Sagaidak, R.N.},
  \bibinfo{author}{Shirokovsky, I.V.}, \bibinfo{author}{Tsyganov, Y.S.},
  \bibinfo{author}{Voinov, A.A.}, \bibinfo{author}{Gulbekian, G.G.},
  \bibinfo{author}{Bogomolov, S.L.}, \bibinfo{author}{Gikal, B.N.},
  \bibinfo{author}{Mezentsev, A.N.}, \bibinfo{author}{Iliev, S.},
  \bibinfo{author}{Subbotin, V.G.}, \bibinfo{author}{Sukhov, A.M.},
  \bibinfo{author}{Subotic, K.}, \bibinfo{author}{Zagrebaev, V.I.},
  \bibinfo{author}{Vostokin, G.K.}, \bibinfo{author}{Itkis, M.G.},
  \bibinfo{author}{Moody, K.J.}, \bibinfo{author}{Patin, J.B.},
  \bibinfo{author}{Shaughnessy, D.A.}, \bibinfo{author}{Stoyer, M.A.},
  \bibinfo{author}{Stoyer, N.J.}, \bibinfo{author}{Wilk, P.A.},
  \bibinfo{author}{Kenneally, J.M.}, \bibinfo{author}{Landrum, J.H.},
  \bibinfo{author}{Wild, J.F.}, \bibinfo{author}{Lougheed, R.W.},
  \bibinfo{year}{2006}.
\newblock \bibinfo{title}{Synthesis of the isotopes of elements 118 and 116 in
  the $^{249}\mathrm{Cf}$ and $^{245}\mathrm{Cm}+^{48}\mathrm{Ca}$ fusion
  reactions}.
\newblock \bibinfo{journal}{Phys. Rev. C} \bibinfo{volume}{74},
  \bibinfo{pages}{044602}.
\newblock \DOIprefix\doi{10.1103/PhysRevC.74.044602}.
\bibitem[{Oganessian et~al.(2010)}]{Oganessian10}
\bibinfo{author}{Oganessian, Y.T.}, et~al., \bibinfo{year}{2010}.
\newblock \bibinfo{title}{Synthesis of a new element with atomic number z =
  117}.
\newblock \bibinfo{journal}{Phys. Rev. Lett.} \bibinfo{volume}{104},
  \bibinfo{pages}{142502}.
\newblock \DOIprefix\doi{10.1103/PhysRevLett.104.142502}.
\bibitem[{Stefanini et~al.(2024)Stefanini, Montagnoli., Fabbro, Corradi,
  Fioretto and Szilner}]{Stefanini24}
\bibinfo{author}{Stefanini, A.M.}, \bibinfo{author}{Montagnoli., G.},
  \bibinfo{author}{Fabbro, M.D.}, \bibinfo{author}{Corradi, L.},
  \bibinfo{author}{Fioretto, E.}, \bibinfo{author}{Szilner, S.},
  \bibinfo{year}{2024}.
\newblock \bibinfo{title}{The slopes of sub-barrier heavy-ion fusion excitation
  functions shed light on the dynamics of quantum tunnelling}.
\newblock \bibinfo{journal}{Sci. Rep.} \bibinfo{volume}{14},
  \bibinfo{pages}{12849}.
\newblock \DOIprefix\doi{10.1038/s41598-024-63107-7}.
\bibitem[{Thompson(1988)}]{THOMPSON1988167}
\bibinfo{author}{Thompson, I.J.}, \bibinfo{year}{1988}.
\newblock \bibinfo{title}{Coupled reaction channels calculations in nuclear
  physics}.
\newblock \bibinfo{journal}{Computer Physics Reports} \bibinfo{volume}{7},
  \bibinfo{pages}{167--212}.
\newblock \DOIprefix\doi{10.1016/0167-7977(88)90005-6}.
\bibitem[{Wiescher et~al.(2025)Wiescher, Bertulani, Brune, deBoer, Diaz-Torres,
  Gasques, Langanke, Navrátil, Nazarewicz, Okołowicz, Phillips, Płoszajczak,
  Quaglioni and Tumino}]{Wiescher25}
\bibinfo{author}{Wiescher, M.}, \bibinfo{author}{Bertulani, C.A.},
  \bibinfo{author}{Brune, C.R.}, \bibinfo{author}{deBoer, R.J.},
  \bibinfo{author}{Diaz-Torres, A.}, \bibinfo{author}{Gasques, L.R.},
  \bibinfo{author}{Langanke, K.}, \bibinfo{author}{Navrátil, P.},
  \bibinfo{author}{Nazarewicz, W.}, \bibinfo{author}{Okołowicz, J.},
  \bibinfo{author}{Phillips, D.R.}, \bibinfo{author}{Płoszajczak, M.},
  \bibinfo{author}{Quaglioni, S.}, \bibinfo{author}{Tumino, A.},
  \bibinfo{year}{2025}.
\newblock \bibinfo{title}{Quantum physics of stars}.
\newblock \bibinfo{journal}{Rev. of Mod. Phys.} \bibinfo{volume}{97},
  \bibinfo{pages}{025003}.
\newblock \DOIprefix\doi{10.1103/RevModPhys.97.025003}.
\bibitem[{Yagi et~al.(1962)Yagi, Katori and Ohnuma}]{Yagi62}
\bibinfo{author}{Yagi, K.}, \bibinfo{author}{Katori, K.},
  \bibinfo{author}{Ohnuma, H.}, \bibinfo{year}{1962}.
\newblock \bibinfo{title}{{E}xperiment on elastic scattering of protons by
  $^{18}$o}.
\newblock \bibinfo{journal}{J. Phys. Soc. Japan} \bibinfo{volume}{17},
  \bibinfo{pages}{595}.
\newblock \DOIprefix\doi{10.1143/JPSJ.17.595}.

\end{thebibliography}






\end{document}